\documentclass[amsmath,amssymb,showkeys, showpacs, superscriptaddress]{revtex4-1}
\usepackage{graphicx}
\usepackage{amsthm}%
\usepackage{amsmath}
\usepackage{amssymb}
\usepackage{phonetic}
\usepackage{color}
\usepackage{mathrsfs}
\usepackage{tikz-cd}
\usetikzlibrary{positioning,arrows,calc}
 \usepackage{cancel}
\expandafter\ifx\csname package@font\endcsname\relax\else
 \expandafter\expandafter
 \expandafter\usepackage
 \expandafter\expandafter
 \expandafter{\csname package@font\endcsname}%
\fi
\newtheorem{theorem}{Theorem}[section]

\theoremstyle{definition}

\parskip \baselineskip

\newcommand{\no}{\noindent}

\begin{document}

\vspace*{0.4 in}
\title{Exact normalized eigenfunctions for general deformed Hulth\'en potentials}
\author{Richard L. Hall}
\affiliation{Department of Mathematics and Statistics, Concordia University,
1455 de Maisonneuve Boulevard West, Montr\'eal,
Qu\'ebec, Canada H3G 1M8,}\email{richard.hall@concordia.ca}
\author{Nasser Saad}
\affiliation{School of Mathematical and Computational Sciences,
University of Prince Edward Island, 550 University Avenue,
Charlottetown, PEI, Canada C1A 4P3,}\email{nsaad@upei.ca}
\author{K. D. Sen}
\affiliation{INSA Senior Scientist, School of Chemistry, University  of Hyderabad 500046, India.}
\email{kds77@uohyd.ac.in}

\begin{abstract}
\noindent  The exact solutions of Schr\"odinger's equation with the deformed Hulth\'en potential
$V_q(x)=-{\mu\, e^{-\delta\,x }}/({1-q\,e^{-\delta\,x}}),~ \delta,\mu, q>0$ are given, along with a closed--form formula for the normalization constants of the eigenfunctions for arbitrary $q>0$. The Crum-Darboux transformation is then used to derive the corresponding exact solutions for the extended Hulth\'en potentials $V(x)=   -{\mu\, e^{-\delta\,x }}/({1-q\,e^{-\delta\,x}})+ {q\,j(j+1)\, e^{-\delta\,x }}/({1-q\,e^{-\delta\,x}})^2, j=0,1,2,\dots.$ A general formula for the new normalization condition is also provided.\end{abstract}

\maketitle
\section{Introduction}\label{intro}
\noindent The Hulth\'en potentials \cite{hu,flugge}
\begin{align}\label{eq1}
V(x)=-\dfrac{\mu\,e^{-\delta\,x }}{1-e^{-\delta\,x}},\qquad x\in (0,\infty),
\end{align}
where $\mu$ is a constant and $\delta>0$ is the screening parameter that
determines the potential range, have important applications in nuclear and particle physics, in atomic physics, and in condensed--matter physics \cite{Hulthen, Durand, van, Gruninger, Szalcwicz,Lindhard}. Except for the $\ell$=0 case, where the exact energy eigenvalues are known, a variety of numerical methods have been employed in the literature to obtain the eigenvalues and eigenfunctions. The so called centrifugal term approximation \cite{Ferr} has been widely used in these calculations. Interesting energy level ordering involving this potential has been derived using the elementary comparison theorem of quantum mechanics which serves to provide  deeper insights \cite{Richard1,Richard2} in to the various other screened Coulomb potentials. A variation of the Hulth\'en potential can be written as
\begin{align}\label{eq2}
V_q(x)=-\dfrac{\mu\, e^{-\delta\,x }}{1-q\,e^{-\delta\,x}}\quad q>0, ~ x\in (\log(q)/\delta,\infty).
\end{align}
Although such potential variations may be found  in the literature \cite{Eriges, Berk,Akcay} for a variety of applications, the potential in \eqref{eq2} is essentially the same as the original potential \eqref{eq1} since a simple shift of the independent variable such as $x\to x+\log(q)/\delta$ transforms $V_q(x)$  back into $V(x)$ with the only difference being a new  coupling constant $\mu$. In future developments, in which angular momentum will be considered, it may be advantageous to revert the viewpoint to \eqref{eq1} so that the potential and the centrifugal terms all have singularities at the origin.  However, since $V_q(x)$ has been widely used in the literature, we shall adopt this form in the present work. Therefore, in atomic units, Schr\"odinger's equation for our problem becomes
\begin{equation}\label{eq3}
-\frac{1}{2} \dfrac{d^2\psi(x)}{dx^2}-\dfrac{\mu\, e^{-\delta\,x} }{1- q\,e^{-\delta\,x} }\,\psi(x)=E\,\psi(x),\quad
\int_{\log(q)/\delta}^\infty |\psi(x)|^2dx=1, \quad \psi(\log(q)/\delta)=\psi(\infty)=0.
\end{equation}
\vskip0.1true in
\noindent To our knowledge, no closed-form expression for the normalization constants for arbitrary $q>0$ was given in the literature (including the classical Hulth\'en potential \eqref{eq1}). The present work provides the complete normalized solutions of Schr\"odinger's equation \eqref{eq3}. Crum-Darboux transformations are then used to generate the eigenfunctions of the equation
 \begin{equation}\label{eq4}
-\frac{1}{2} \dfrac{d^2\phi(x)}{dx^2}+\left(\dfrac{\mu\, e^{-\delta\,x} }{(1- q\,e^{-\delta\,x})^2 } -\dfrac{v\, e^{-\delta\,x} }{1- q\,e^{-\delta\,x} }\right)\,\phi(x)=\mathfrak E\,\phi(x),\quad
\int_{\log(q)/\delta}^\infty |\phi(x)|^2dx=1, \quad \phi(\log(q)/\delta)=\phi(\infty)=0.
\end{equation}
A general formula is provided for the normalization constants of the eigenfunctions $\phi(x)$ in terms of the eigenfunctions $\psi(x)$, for arbitrary $q>0$. 
\vskip0.1true in
\noindent The paper is organized as follows: in section~II, we discuss the exact solutions of  Schr\"odinger's equation \eqref{eq3}. In section~III, we develop an analytic expression for the normalization constants in terms of the generalized hypergeometric function ${}_3F_2$. In section~IV, we give a general review of the Crum-Darboux transformation, and a simplified formula for the normalization constants. These ideas are used in section~V to generate the exact eigenfunctions of equation \eqref{eq4}. The approach given here provides the logical framework \cite{Dong2007,Dong2007a,Dong2007b,Dong2008} behind the success of the 
centrifugal--term approximation $1/x^2\approx e^{-\delta x}/(1-e^{-\delta x})^2$ used to estimate the eigenvalues and eigenfunctions for the Hulth\'en potential ($q=1$) for $\ell\neq 0$. 
\section{Generalized Hulth\'en potential: exact solutions}\label{sec2}
\noindent In this section, we show that the exact solutions of Schr\"odinger's equation \eqref{eq3} may be  expressed in terms of the Gauss hypergeometric functions as
\begin{equation}\label{eq5}
\psi(x)=e^{-\sqrt{- 2\,E}\,x}(1-q\,e^{-\delta\, x})\,{} _2F_1\left(1+\frac{\sqrt{-2\, E}}{\delta}-\frac{1}{\delta}\sqrt{\frac{2\mu}{q}-2 E},1+\frac{\sqrt{-2\, E}}{\delta}+
\frac{1}{\delta}\sqrt{\frac{2\mu}{q}-2 E};1+\frac{2\sqrt{-2\, E}}{\delta};q\, e^{-\delta\,x}\right).
\end{equation}
Indeed, the change of variable $z=e^{-\delta\,x}$ allows equation \eqref{eq3} to be written as
\begin{equation}\label{eq6}
z^2\dfrac{d^2\psi}{dz^2}+z\dfrac{d\psi}{dz}+\frac{v\,z}{1-q\,z}\psi=-\varepsilon\,\psi,\quad 0< z< \frac1q,\quad
\psi(0)=\psi\left(1/q\right)=0,
\end{equation}
where
$v= {2\mu}/\delta^2$ and $\varepsilon= 2 E/\delta^2.$ The differential equation \eqref{eq6} has two regular singular points $z=0$ and $z=1/q$ with the indicial equations $\eta^2+ \varepsilon=0$ (or $\eta=\sqrt{-2 E}/\delta$) and $s(s-1)=0$ (or $s=0,1$), respectively, in addition to an irregular singular point at $z=\infty$. Thus, the general solution of equation \eqref{eq6} assumes the form
 \begin{equation}\label{eq7}
 \psi(z)=z^\eta\left(1-q z\right)^s f(z).
 \end{equation}
 The factor $(1-q z)^{s}$ must have $s$ a positive integer so that $\psi(1/q) = 0$.  Thus $s$ cannot be zero.   On substituting the ansatz \eqref{eq7} in equation \eqref{eq6}, it is not difficult to show that the function $f(z)$, after implementing the indicial equations, satisfies the differential equation
 \begin{align}\label{eq8}
z(1-q\,z)f''(z)&+(-q(2 \eta+2 s+1)\,z+2 \eta+1)f'(z)+\left(v-q\,s\, \left(2\,\eta+s\right)\right)f(z)=0.
\end{align}
This is a hypergeometric differential equation with exact solutions given, in terms of the Gauss hypergeometric functions, as
 \begin{align}\label{eq9}
f(z)&=C_1 \, _2F_1\left(\eta+s-\sqrt{\eta^2 +\frac{v}{q}},\eta+s+
\sqrt{\eta^2+\frac{v}{q}};1+2\, \eta;q\, z\right)\notag\\
&+C_2\, z^{-2 \eta} \, _2F_1\left(s-\eta-\sqrt{\eta^2+\frac{v}{q}},
s-\eta+\sqrt{\eta^2+\frac{v}{q}};1-2\, \eta;q\, z\right).
\end{align}
However, the boundary condition $\psi(0)=0$ forces that the vanishing of the constant $C_2=0$. Thus,
 \begin{align}\label{eq10}
f(z)&={} _2F_1\left(\eta+s-\sqrt{\eta^2 +\frac{v}{q}},\eta+s+
\sqrt{\eta^2+\frac{v}{q}};2\, \eta+1;q\, z\right),
\end{align}
from which the general solution of Eq. \eqref{eq3} takes the form \eqref{eq7}. For polynomial solutions $f_n(z)$, the termination of the hypergeometric series \eqref{eq10} requires
$
 \eta+1-\sqrt{\eta^2 +\frac{v}{q}}=-n,~ n=0,1,2,\dots.
$ that yields the following expression for the eigenvalues (using $\eta=\sqrt{-2E_n}/\delta$, $v=2\mu/\delta^2$ and $s=1$)
 \begin{align}\label{eq11}
 E_n&= -\frac{1}{2}\left(\frac{\mu}{q\,\delta\,(1+n)}-\frac{\delta\,(1+n)}{2}\right)^2,
 \end{align}
 with the exact (unnormalized) wave functions:
 \begin{align}\label{eq12}
 \psi_n(x)&=N_n\left(1-q\, e^{-\delta\, x}\right) e^{-\left(\frac{\mu}{q\,\delta\, (1+n)} -\frac{(n+1)\delta}{2}\right)\,x}\,{}_2F_1\left(-n,1+\frac{2 \mu}{q\,\delta^2 (1+n)};\frac{2 \mu}{q\,\delta^2\, (1+n)}-n;q\,e^{-\delta\, x}\right)
 \end{align}
 up to normalization constant $N_n$. Note, by using the Pfaff transformation identity
 \begin{align}\label{eq13}
 {}_2F_1(\alpha,\beta;\gamma;z)=(1-z)^{\gamma-\alpha-\beta}{}_2F_1(\gamma-\alpha,\gamma-\beta;\gamma;z),
 \end{align}
 we can write the exact solution \eqref{eq12} as
 \begin{align}\label{eq14}
 \psi_n(x)&=N_ne^{-\left(\frac{\mu}{q\,\delta\, (1+n)} -\frac{(n+1)\delta}{2}\right)\,x}\,{}_2F_1\left(-n-1,\frac{2 \mu}{q\,\delta^2 (1+n)};\frac{2 \mu}{q\,\delta^2\, (1+n)}-n;q\,e^{-\delta x}\right).
 \end{align}
 Thus, the number of the discrete bound-states is bounded above by the inequality
 \begin{align}\label{eq15}
 0\leq n<-1+ \frac{1}{\delta}\sqrt{\frac{2\,\mu}{q}} ,\qquad where\qquad 0<q<\frac{2\, \mu}{\delta^2}.
 \end{align}
\section{Generalized Hulth\'en potential: normalization constant}\label{sec3}
 \noindent The normalization constant $N_n$ in equation \eqref{eq12} can be evaluated, for $2\mu>q\,\delta^2 (m+1) (n+1)$, using the following definite integral
\begin{align}\label{eq16}
I_{nm}=\int_{\log(q)/\delta}^\infty e^{-\frac{(2+m+n) \left(2 \mu-q\,\delta^2 (1+m) (1+n)\right) x}{2 \delta (1+m) (1+n) q}}& \left(1-q\,e^{-\delta x}\right)^2\, _2F_1\left(-n,\frac{2 \mu}{\delta^2 (n+1) q}+1;\frac{2 \mu}{\delta^2 (n+1) q}-n;q e^{-\delta\, x}\right)\notag\\
&\times \, _2F_1\left(-m,\frac{2 \mu}{\delta^2 (m+1) q}+1;\frac{2 \mu}{\delta^2 (m+1) q}-m;q e^{-\delta\, x}\right)dx.
\end{align}
To evaluate this definite integral, we use the change of variables $\tau=q\, e^{-\delta\,x}$ and note for $x=\log(q)/\delta,$ that $\tau=1/q$, while for $x=\infty, \tau=0$. Further, by the series representation of the Hypergeometric function, it follows that
\begin{align}\label{eq17}
I_{nm}=\frac{1}{\delta}
\sum_{i=0}^n
\sum_{j=0}^m
\dfrac{(-n)_i\left(1+\frac{2\mu}{q\delta^2(1+n)}\right)_i}{\left(\frac{2\mu}{q\delta^2(1+n)}-n\right)_i\,i!}\dfrac{(-m)_j\left(1+\frac{2\mu}{q\delta^2(1+m)}\right)_j}{\left(\frac{2\mu}{q\delta^2(1+m)}-m\right)_j\,j!}q^{i+j}\int_0^{1/q}(q \tau-1)^2 \tau^{\frac{\mu(m+n+2)}{\delta^2 (m+1) (n+1) q}-\frac{m}{2}-\frac{n}{2}-2+i+j}d\tau.
\end{align}
The integral on the left-hand side can be evaluated in terms of the Gamma function to yield
\begin{align}\label{eq18}
I_{nm}=&\frac{2 q^{1+\frac{m+n}{2}-\frac{\mu (m+n+2)}{\delta^2 (m+1) (n+1) q}} \Gamma \left(\frac{\mu (m+n+2)}{\delta^2 (m+1) (n+1) q}-\frac{m+n}{2}-1\right)}{\delta\, \Gamma \left(2-\frac{m+n}{2}+\frac{\mu (m+n+2)}{\delta^2 (m+1) (n+1) q}\right)}\notag\\
&\times\sum _{i=0}^n \sum _{j=0}^m \frac{(-n)_i (-m)_j \left(\frac{2 \mu}{q \delta^2 (n+1)}+1\right)_i \left(\frac{2 \mu}{q \delta^2 (m+1)}+1\right)_j \left(\frac{\mu (m+n+2)}{\delta^2 (m+1) (n+1) q}-\frac{m+n}{2}-1\right)_{i+j}}{i!\, j!\, \left(\frac{2 \mu}{q \delta^2 (n+1)}-n\right)_i \left(\frac{2\mu}{q\delta^2 (m+1)}-m\right)_j\left(\frac{\mu (m+n+2)}{\delta^2 (m+1) (n+1) q}-\frac{m+n}{2}+2\right)_{i+j}}\end{align}
Using the Pochhammer identity $(\alpha)_{i+j}=(\alpha+i)_j(\alpha)_i\Gamma(\alpha)$, equation \eqref{eq18} can be written as
\begin{align}\label{eq19}
I_{nm}&=\frac{2 q^{1+\frac{m+n}{2}-\frac{\mu (m+n+2)}{q\,\delta^2 (m+1) (n+1)}} \Gamma \left(\frac{(m+n+2) \mu}{\delta^2 (m+1) (n+1) q}-\frac{m+n}{2}-1\right)}{\delta \,\Gamma \left(\frac{(m+n+2) \mu}{\delta^2 (m+1) (n+1) q}-\frac{m+n}{2}+2\right)}\sum _{i=0}^n \frac{(-n)_i \left(\frac{2 \mu}{q \delta^2 (n+1)}+1\right)_i \left(\frac{(m+n+2) \mu}{\delta^2 (m+1) (n+1) q}-\frac{m+n}{2}-1\right)_i}{i! \left(\frac{2 \mu}{q \delta^2 (n+1)}-n\right)_i \left(\frac{(m+n+2) \mu}{q\,\delta^2 (m+1) (n+1)}-\frac{m+n}{2}+2\right)_i}\notag\\
&\times {}_3F_2\left(\begin{array}{lll}-m,&\frac{2 \mu}{q\,\delta^2 (1+m)}+1,&i-1-\frac{m+n}{2}+\frac{\mu\,(2+m+n)}{q\,\delta^2 (1+m) (1+n)}\\
\frac{2 \mu}{q\,\delta^2(1+m)}-m,&i+2-\frac{m+n}{2}+\frac{\mu\, (2+m+n)}{q\,\delta^2 (1+m) (1+n)}&~\\
\end{array}\bigg|1\right)\end{align}
For $m=n$, it follow that
\begin{align}\label{eq20}
I_{nn}&=\frac{2 q^{1+n-\frac{2\mu}{q\,\delta^2 (n+1)}} \Gamma \left(\frac{2\mu}{\delta^2 (n+1) q}-n-1\right)}{\delta \,\Gamma \left(\frac{2 \mu}{\delta^2 (n+1) q}-n+2\right)}\sum _{i=0}^n \frac{(-n)_i \left(\frac{2 \mu}{q \delta^2 (n+1)}+1\right)_i \left(\frac{2\mu}{\delta^2  (n+1) q}-n-1\right)_i}{i! \left(\frac{2 \mu}{q \delta^2 (n+1)}-n\right)_i \left(\frac{2 \mu}{q\,\delta^2 (n+1)}-n+2\right)_i}\notag\\
&\times{}_3F_2\left(\begin{array}{lll}-n,&\frac{2 \mu}{q\,\delta^2 (1+n)}+1,&i+\frac{2\mu}{q\,\delta^2 (1+n)}-n-1\\
\frac{2 \mu}{q\,\delta^2(1+n)}-n,&i+2-n+\frac{2\mu}{q\,\delta^2 (1+n)}&~\\
\end{array}\bigg|1\right).
\end{align}
Breaking the finite sum, the equation can be written as
\begin{align}\label{eq21}
I_{nn}&=\frac{2 q^{1+n-\frac{2 \mu}{q\,\delta^2 (n+1)}} \Gamma \left(\frac{2 \mu}{q\,\delta^2 (n+1)}-n-1\right)}{\delta\, \Gamma \left(\frac{2 \mu}{q\,\delta^2 (n+1)}-n+2\right)}\, \,{}_3F_2\left(\begin{array}{lll}-n,&\frac{2 \mu}{q\,\delta^2 (n+1)}+1,&\frac{2 \mu}{q\,\delta^2 (n+1)}-n-1\\
\frac{2 \mu}{\delta^2 (n+1) q}-n,&2-n+\frac{2 \mu}{q\,\delta^2 (n+1)},& \end{array}\bigg|1\right)\notag\\
&+\frac{2 q^{1+n-\frac{2\mu}{q\,\delta^2 (n+1)}} \Gamma \left(\frac{2\mu}{\delta^2 (n+1) q}-n-1\right)}{\delta \,\Gamma \left(\frac{2 \mu}{\delta^2 (n+1) q}-n+2\right)}\sum _{i=1}^n \frac{(-n)_i \left(\frac{2 \mu}{q \delta^2 (n+1)}+1\right)_i \left(\frac{2\mu}{\delta^2  (n+1) q}-n-1\right)_i}{i! \left(\frac{2 \mu}{q \delta^2 (n+1)}-n\right)_i \left(\frac{2 \mu}{q\,\delta^2 (n+1)}-n+2\right)_i}\notag\\
&\times{}_3F_2\left(\begin{array}{lll}-n,&\frac{2 \mu}{q\,\delta^2 (1+n)}+1,&i+\frac{2\mu}{q\,\delta^2 (1+n)}-n-1\\
\frac{2 \mu}{q\,\delta^2(1+n)}-n,&i+2-n+\frac{2\mu}{q\,\delta^2 (1+n)}&~\\
\end{array}\bigg|1\right)
\end{align}
Using the hypergeometric identity \cite[formula 7.4.4.1]{SpFu}
\begin{align}\label{eq22}
{}_3F_2\left(\begin{array}{lll}a,&b,&c\\ d,&e,& \end{array}\bigg|1\right)&=\frac{\Gamma(d)\Gamma(d+e-a-b-c)}{\Gamma(d+e-a-b)\Gamma(d-c)}
{}_3F_2\left(\begin{array}{lll}e-a,&e-b,&c\\ d+e-a-b,&e,& \end{array}\bigg|1\right),\notag\\
&\hskip1.5true in (Re(d+e-a-b-c)>0,~Re(d-c)>0),
\end{align}
it follows
\begin{align*}
I_{nn}&=\frac{2 q^{1+n-\frac{2 \mu}{q\,\delta^2 (n+1)}} \Gamma \left(\frac{2 \mu}{q\,\delta^2 (n+1)}-n-1\right)}{\delta\, \Gamma \left(\frac{2 \mu}{q\,\delta^2 (n+1)}-n+2\right)}\, \,{}_3F_2\left(\begin{array}{lll}-n,&\frac{2 \mu}{q\,\delta^2 (n+1)}+1,&\frac{2 \mu}{q\,\delta^2 (n+1)}-n-1\\
\frac{2 \mu}{\delta^2 (n+1) q}-n,&2-n+\frac{2 \mu}{q\,\delta^2 (n+1)},& \end{array}\bigg|1\right)\notag\\
&+\frac{2 q^{1+n-\frac{2\mu}{q\,\delta^2 (n+1)}} \Gamma \left(\frac{2\mu}{\delta^2 (n+1) q}-n-1\right)}{\delta \,\Gamma \left(\frac{2 \mu}{\delta^2 (n+1) q}-n+2\right)}
\sum _{i=1}^n \frac{(-n)_i \left(\frac{2 \mu}{\delta^2 (n+1) q}+1\right)_i \left(\frac{(2 n+2) \mu}{\delta^2 (n+1)^2 q}-n-1\right)_i \Gamma \left(i-n+\frac{2 \mu}{\delta^2 (n+1) q}+2\right)}{\,i! \left(\frac{2 \mu}{q\,\delta^2 (n+1)}-n\right)_i \left(\frac{(2 n+2) \mu}{q\,\delta^2 (n+1)^2}-n+2\right)_i\Gamma (i-n+1)\, \Gamma \left(\frac{2 \mu}{q\,\delta^2 (n+1)}+3\right)}\notag\\
&\times\, _3F_2\left(1-i,\frac{2 \mu}{\delta^2 (n+1) q}+1,\frac{2 \mu}{q\,\delta^2 (n+1)};\frac{2 \mu}{q\,\delta^2 (n+1)}+3,\frac{2 \mu}{q\,\delta^2 (n+1)}-n;1\right)
\end{align*}
However, because of the reciprocal of the Gamma function $1/\Gamma(i-n+1)$, where $1/\Gamma(-m)=0,m=0,1,2,\dots$, the finite sum survives only for $i=n$ (otherwise each term is zero) whence
\begin{align}\label{eq23}
I_{nn}&=\frac{2 q^{1+n-\frac{2 \mu}{q\,\delta^2 (n+1)}} \Gamma \left(\frac{2 \mu}{q\,\delta^2 (n+1)}-n-1\right)}{\delta\, \Gamma \left(\frac{2 \mu}{q\,\delta^2 (n+1)}-n+2\right)}\, \,{}_3F_2\left(\begin{array}{lll}-n,&\frac{2 \mu}{q\,\delta^2 (n+1)}+1,&\frac{2 \mu}{q\,\delta^2 (n+1)}-n-1\\
\frac{2 \mu}{\delta^2 (n+1) q}-n,&2-n+\frac{2 \mu}{q\,\delta^2 (n+1)},& \end{array}\bigg|1\right)\notag\\
&+\frac{2 q^{1+n-\frac{2\mu}{q\,\delta^2 (n+1)}} \Gamma \left(\frac{2\mu}{\delta^2 (n+1) q}-n-1\right)}{\delta \,\Gamma \left(\frac{2 \mu}{\delta^2 (n+1) q}-n+2\right)}
\frac{(-n)_n \left(\frac{2 \mu}{q\,\delta^2 (n+1)}+1\right)_n \left(\frac{2\mu}{q\,\delta^2 (n+1)}-n-1\right)_n \Gamma \left(\frac{2 \mu}{q\,\delta^2 (n+1)}+2\right)}{\,n! \left(\frac{2 \mu}{q\,\delta^2 (n+1)}-n\right)_n \left(\frac{(2 n+2) \mu}{q\,\delta^2 (n+1)^2}-n+2\right)_n\, \Gamma \left(\frac{2 \mu}{q\,\delta^2 (n+1)}+3\right)}\notag\\
&\times\, _3F_2\left(1-n,\frac{2 \mu}{\delta^2 (n+1) q}+1,\frac{2 \mu}{q\,\delta^2 (n+1)};\frac{2 \mu}{q\,\delta^2 (n+1)}+3,\frac{2 \mu}{q\,\delta^2 (n+1)}-n;1\right).
\end{align}
Upon using the Pochhammer identities
$$(-n)_n=(-1)^n\, n!,\quad\frac{\left(\frac{2 \mu}{q\,\delta^2 (n+1)}-n-1\right)_n}{\left(\frac{2 \mu}{\delta^2 (n+1) q}-n\right)_n}=\frac{\frac{2 \mu}{q\,\delta^2 (n+1)}-n-1}{\frac{2 \mu}{q\,\delta^2 (n+1) }-1},\quad \frac{\left(\frac{2 \mu}{\delta^2 (n+1) q}+1\right)_n}{\left(\frac{2 \mu}{q\,\delta^2 (n+1)}-n+2\right)_n}=\frac{(-1)^n \left(\frac{2 \mu}{q\,\delta^2 (n+1)}+1\right)_n}{\left(-\frac{2 \mu}{q\,\delta^2 (n+1)}-1\right){}_n},$$
it follows finally that
\begin{align}\label{eq24}
&\int _{{\log (q)/\delta}}^{\infty}\left(1-q\,e^{-\delta\,x}\right)^2 e^{-\left(\frac{2 \mu}{\delta\, q\,(1+n)}-(1+n)\delta\right) x}\left[{}_2F_1\left(-n,\frac{2 \mu}{q\,\delta^2 (n+1)}+1;\frac{2 \mu}{q\,\delta^2 (n+1)}-n;q\,e^{-\delta x}\right)\right]^2dx\notag\\
&=\frac{2 q^{n+1-\frac{2 \mu}{q\,\delta^2 (n+1)}}\, \Gamma \left(\frac{2 \mu}{q\,\delta^2 (n+1)}-n-1\right)}{\delta\, \Gamma \left(\frac{2\mu}{q\,\delta^2 (n+1)}-n+2\right)}\bigg[\frac{q\,\delta^2 (n+1)\left(2 \mu-q\,\delta^2 (n+1)^2\right) \left(\frac{2\mu}{q\,\delta^2 (n+1)}+1\right)_n}{2 \left(2 \mu-q\,\delta^2 (n+1)\right) \left(q\,\delta^2 (n+1)+\mu\right) \left(-\frac{2\mu}{q\,\delta^2 (n+1)}-1\right)_n}\notag\\
&\times{}_3F_2\left(\begin{array}{lll}1-n,&\frac{2 \mu}{q\,\delta^2 (n+1)}+1,&\frac{2 \mu}{q\,\delta^2 (n+1)}\\
\frac{2\mu}{q\,\delta^2 (n+1)}+3,&\frac{2 \mu}{q\,\delta^2 (n+1)}-n,& \end{array}\bigg|1\right)+\, _3F_2\left(\begin{array}{lll}-n,&\frac{2\mu}{q\,\delta^2 (n+1)}+1,&\frac{2 \mu}{q\,\delta^2 (n+1)}-n-1\\
\frac{2 \mu}{q\,\delta^2 (n+1)}-n,&\frac{2 \mu}{q\,\delta^2 (n+1)}-n+2& \end{array}\bigg|1\right)\bigg].
\end{align}
Next, for the case where $m\neq n$,
\begin{align}\label{eq25}
I_{nm}&=\frac{2 q^{1+\frac{m+n}{2}-\frac{\mu (m+n+2)}{q\,\delta^2 (m+1) (n+1)}} \Gamma \left(\frac{(m+n+2) \mu}{\delta^2 (m+1) (n+1) q}-\frac{m+n}{2}-1\right)}{\delta \,\Gamma \left(\frac{(m+n+2) \mu}{\delta^2 (m+1) (n+1) q}-\frac{m+n}{2}+2\right)}\sum _{i=0}^n \frac{(-n)_i \left(\frac{2 \mu}{q \delta^2 (n+1)}+1\right)_i \left(\frac{(m+n+2) \mu}{\delta^2 (m+1) (n+1) q}-\frac{m+n}{2}-1\right)_i}{i! \left(\frac{2 \mu}{q \delta^2 (n+1)}-n\right)_i \left(\frac{(m+n+2) \mu}{q\,\delta^2 (m+1) (n+1)}-\frac{m+n}{2}+2\right)_i}\notag\\
&\times {}_3F_2\left(\begin{array}{lll}-m,&\frac{2 \mu}{q\,\delta^2 (1+m)}+1,&i-1-\frac{m+n}{2}+\frac{\mu\,(2+m+n)}{q\,\delta^2 (1+m) (1+n)}\\
\frac{2 \mu}{q\,\delta^2(1+m)}-m,&i+2-\frac{m+n}{2}+\frac{\mu\, (2+m+n)}{q\,\delta^2 (1+m) (1+n)}&~\\
\end{array}\bigg|1\right).
\end{align}
Using the identity \cite[formula 7.4.4.90]{SpFu}
\begin{align}\label{eq26}
{}_3F_2\left(\begin{array}{lll}-m,&a,&b\\ a-\ell,& b-s,& \end{array}\bigg| 1\right)=0,\qquad {\rm if}\quad  (\ell+s=1,2,3,\dots,m-1),
\end{align}
we see that it is enough to consider the case $n=0$ and $m\neq 0$, for every other (fixed) value of $n$, the parameter $a$ is varied by a constant factor. Thus,
\begin{align}\label{eq27}
I_{0m}&=\frac{2 q^{1+\frac{m}{2}-\frac{\mu (m+2)}{q\,\delta^2 (m+1)}} \Gamma \left(\frac{(m+2) \mu}{q\,\delta^2 (m+1)}-\frac{m}{2}-1\right)}{\delta \,\Gamma \left(\frac{(m+2) \mu}{q\, \delta^2 (m+1)}-\frac{m}{2}+2\right)}  {}_3F_2\left(\begin{array}{lll}-m,&\frac{2 \mu}{q\,\delta^2 (1+m)}+1,&-1-\frac{m}{2}+\frac{\mu\,(2+m)}{q\,\delta^2 (1+m)}\\
\frac{2 \mu}{q\,\delta^2(1+m)}-m,&2-\frac{m}{2}+\frac{\mu\, (2+m)}{q\,\delta^2 (1+m)}&~\\
\end{array}\bigg|1\right)
\end{align}
comparing the equation \eqref{eq27} with equation \eqref{eq26}, we note $\ell = m+1,s=-3$, thus $m-2<m-1$, it follows that
$I_{0m}=0$, noting that by a similar argument, for  $n\neq 0$ and $n\neq m$, each term of the finite sum vanishes . Finally we have
\begin{align}\label{eq28}
&\int _{{\log (q)/\delta}}^{\infty}\left(1-q\,e^{-\delta\,x}\right)^2 e^{-\left(\frac{2 \mu}{\delta\, q\,(1+n)}-(1+n)\delta\right) x}\left[{}_2F_1\left(-n,\frac{2 \mu}{q\,\delta^2 (n+1)}+1;\frac{2 \mu}{q\,\delta^2 (n+1)}-n;q\,e^{-\delta x}\right)\right]^2dx=I_{nm}\delta_{nm},
\end{align}
where
\begin{align}\label{eq29}
I_{nn}&=
\frac{2 q^{n+1-\frac{2 \mu}{q\,\delta^2 (n+1)}}\, \Gamma \left(\frac{2 \mu}{q\,\delta^2 (n+1)}-n-1\right)}{\delta\, \Gamma \left(\frac{2\mu}{q\,\delta^2 (n+1)}-n+2\right)}\bigg[\frac{q\,\delta^2 (n+1)\left(2 \mu-q\,\delta^2 (n+1)^2\right) \left(\frac{2\mu}{q\,\delta^2 (n+1)}+1\right)_n}{2 \left(2 \mu-q\,\delta^2 (n+1)\right) \left(q\,\delta^2 (n+1)+\mu\right) \left(-\frac{2\mu}{q\,\delta^2 (n+1)}-1\right)_n}\notag\\
&\times{}_3F_2\left(\begin{array}{lll}1-n,&\frac{2 \mu}{q\,\delta^2 (n+1)}+1,&\frac{2 \mu}{q\,\delta^2 (n+1)}\\
\frac{2\mu}{q\,\delta^2 (n+1)}+3,&\frac{2 \mu}{q\,\delta^2 (n+1)}-n,& \end{array}\bigg|1\right)+\, _3F_2\left(\begin{array}{lll}-n,&\frac{2\mu}{q\,\delta^2 (n+1)}+1,&\frac{2 \mu}{q\,\delta^2 (n+1)}-n-1\\
\frac{2 \mu}{q\,\delta^2 (n+1)}-n,&\frac{2 \mu}{q\,\delta^2 (n+1)}-n+2& \end{array}\bigg|1\right)\bigg].
\end{align}

\section{Crum-Darboux transformation: Sequential transformations}
\noindent An important technique for generating classes of exactly-solvable quantum potentials is build on the concept of intertwining operators \cite{lange}.  Two  Hamiltonian operators  $\mathcal H_0$ and $\mathcal H_1$ are said to be intertwined if there exist an operator $\mathfrak L$ so that
\begin{equation}\label{eq30}
\mathcal H_1\mathfrak L=\mathfrak L\,\mathcal H_0.
\end{equation}
In this case, if $\phi_{0;n}(x)$ and $\psi_{1;n}(x)$ are eigenfunctions of the intertwined operators  $\mathcal H_0$ and $\mathcal H_1$ respectively, then the two sets of eigenfunctions are related  (\cite{arlen}, p. 63) by the operator $\mathfrak L$ through the relations
\begin{equation}\label{eq31}
\psi_{1;n}(x)=\mathfrak L\, \phi_{0;n}(x),\qquad  \phi_{0;n}(x)=\mathfrak L^\dagger\, \psi_{0;n}(x),
\end{equation}
and the Hamiltonians $\mathcal H_0$ and $\mathcal H_1$ are is said to be isospectral, i.e. share the same spectrum, except for those states that are annihilated by $\mathfrak L$ or $\mathfrak L^\dagger$.  In the context of  the one-dimensional quantum mechanics, $\mathfrak L$ is taken to be a first-order linear differential operator $\mathfrak L=\partial_x+f(x)$ intertwines two one-dimensional Schr\"odinger Hamiltonians $\mathcal H_0=-\partial_{xx}+V(x)$  and $\mathcal H_1=-\partial_{xx}+\mathcal V(x)$ where $x\in (a,b)$ that can be finite or infinite interval. Here, $\partial_x$ refers to the first-derivative with respect to the variable $x$.  Direct computation, using \eqref{eq30}, yields
\begin{align}\label{eq32}
\left(-2\partial_x f(x) +\mathcal V(x)-V(x)\right)\partial_x\phi_n(x) &=\left(\partial_x V(x)+[V(x)-\mathcal V(x)]f(x)+\partial_x^2 f(x)\right)\phi_n(x).
\end{align}
The consistency condition of \eqref{eq32} then requires
\begin{align}\label{eq33}
-2\partial_x f(x)+\mathcal V(x)-V(x)&=0,\quad\mbox{and}\quad \partial_x V(x)+[V(x)-\mathcal V(x)]f(x)+\partial_x^2 f(x)=0.
\end{align}
Substituting the first condition into the second one gives
\begin{align}\label{eq34}
\partial_x\left(V(x)-f^2(x)+\partial_x f(x)\right)&=0\qquad \mbox{or}\qquad
V(x)-f^2(x)+\partial_x f(x)=\lambda
\end{align}
for some constant $\lambda$. Clearly, the function  $f(x)=-\partial_x \phi(x)/\phi(x)=-\partial_x \log\phi(x)$ is a particular solution of the Riccati equation \eqref{eq34} provided that
\begin{align}\label{eq35}
-\partial_{xx}\,\phi(x)+V(x)\phi(x)&=\lambda\,\phi(x),
\end{align}
 that is to say, provided that $\phi$ is an  eigenfunction of the Hamiltonian $\mathcal H_0$ with eigenvalue $\lambda$. An important conclusion from this approach is that every  \emph{no-node} eigenfunction, called a seed function, $\phi=\phi_{0,0}$ of $\mathcal H_0$ (regardless of the normalizability) generates  a new solvable
 Hamiltonian $\mathcal H_1$ with the potential $\mathcal V$ expressed in terms of $V$ and the seed function as
\begin{align}\label{eq36}
\mathcal V(x)=V(x)+2\dfrac{d^2}{dx^2}\log\phi_{0,0}(x).
\end{align}
The first-order differential intertwining operator
\begin{equation}\label{eq37}
\mathfrak L=\dfrac{d}{dx} -\dfrac{1}{\phi_{0,0}}\dfrac{d\phi_{0,0}}{dx}
\end{equation}
 is known as a Darboux transformation \cite{darboux1882}. The following theorem summarize these results:
\begin{theorem}\label{Thm1}
The eigenfunctions $\psi_{1,n}(x),~n=1,2,\dots$ of the Schr\"odinger equation
\begin{align}\label{eq38}
\left(-\dfrac{d^2}{dx^2}+\mathcal V(x)\right)\phi_{1,n}(x)=E_{n}\,\phi_{1,n}(x)\quad\mbox{where}\quad  \mathcal V(x)=V(x)-2\dfrac{d^2}{dx^2}\log \phi_{0,0}(x),\quad x\in(a,b)
\end{align}
are generated using
\begin{align}\label{eq39}
\phi_{1,n}(x)=\left(\dfrac{d}{dx}-\dfrac{\phi_{0,0}'(x)}{\phi_{0,0}(x)}\right) \phi_{0,n}(x)=\dfrac{W(\phi_{0,0}(x),\phi_{0,n}(x))}{W(\phi_{0,0}(x))},\quad n=1,2,\dots,
\end{align}
where $W(\psi_{0,0}(x))\equiv \phi_{0,0}(x)$ and $W(\phi_{0,0}(x),\phi_{0,n}(x))=\psi_{0,0}(x)\psi_{0,n}'(x)-\psi_{0,n}(x)\psi_{0,0}'(x)$ is the classical Wronskian.  Here, $\phi_{0,n}(x), ~n=0,1,2,\dots,$ are solutions of the Schr\"odinger equation
\begin{align}\label{eq40}
-\phi_{0,n}''(x)+ V(x)\phi_{0,n}(x)=E_n\,\phi_{0,n}(x),~n=0,1,2,\dots,~\phi_{0,n}(a)=\phi_{0,n}(b)=0,~ \phi_{0,0}(x)\neq 0~\forall x\in(a,b).
\end{align}
Further, if $\phi_{0,n}(x)$ are normalized according to the condition
\begin{align}\label{eq41}
\int_a^b {\phi_{0,n}(x)}\phi_{0,m}(x)dx=\eta_n\,\delta_{mn},
\end{align}
where $\delta_{mn}$ is the known Kronecker delta, then
\begin{align}\label{eq42}
\int_a^b {\phi_{1,n}(x)}\phi_{1,m}(x)dx=(E_n-E_0)\,\eta_n\,\delta_{mn},\quad n=1,2,\dots.
\end{align}
\end{theorem}
\noindent The proof of the normalization relation \eqref{eq42} is given in the Appendix.  It is evident that such uses of Darboux's transformation can be applied to \eqref{eq39} using
$\mathfrak L_1=\frac{d}{dx}-\frac{\phi_{1,1}'(x)}{\phi_{1,1}}$,
again to reproduce a new solvable eigenvalue problem $\mathcal H_2$ and such a procedure may be repeated an arbitrary number of times \cite{matveev}
as long as the consecutive (generated) Hamiltonians support the existence of (discrete) states, as illustrated by the following diagram:
\begin{center}
\begin{center}
\includegraphics[height=1in,width=4in]{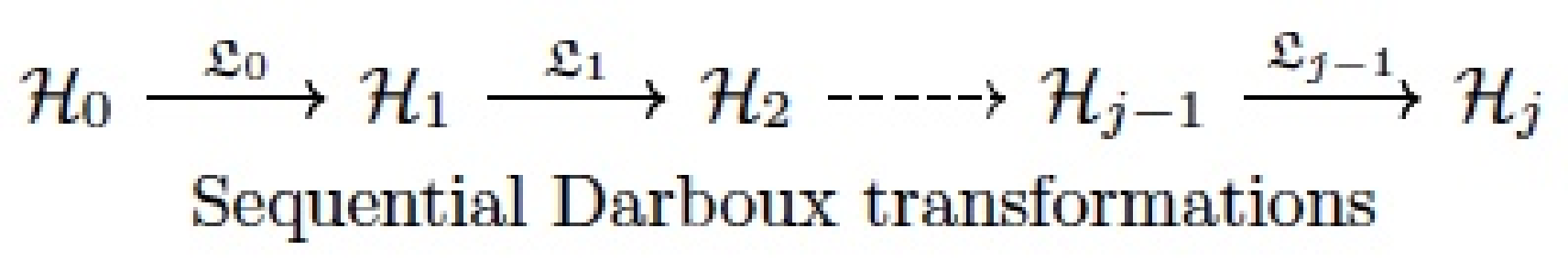}
\end{center}

\end{center}
M. M. Crum \cite{crum} in his remarkable paper 1955 introduced an elegant approach to evaluate the eigenfunctions of the Hamiltonian $\mathcal H_j$,  expressed entirely in terms of the eigenfunctions of the initial Hamiltonian $\mathcal H_0$, without any reference to the intermediate Hamiltonians:
\begin{center}
\includegraphics[height=2in,width=5in]{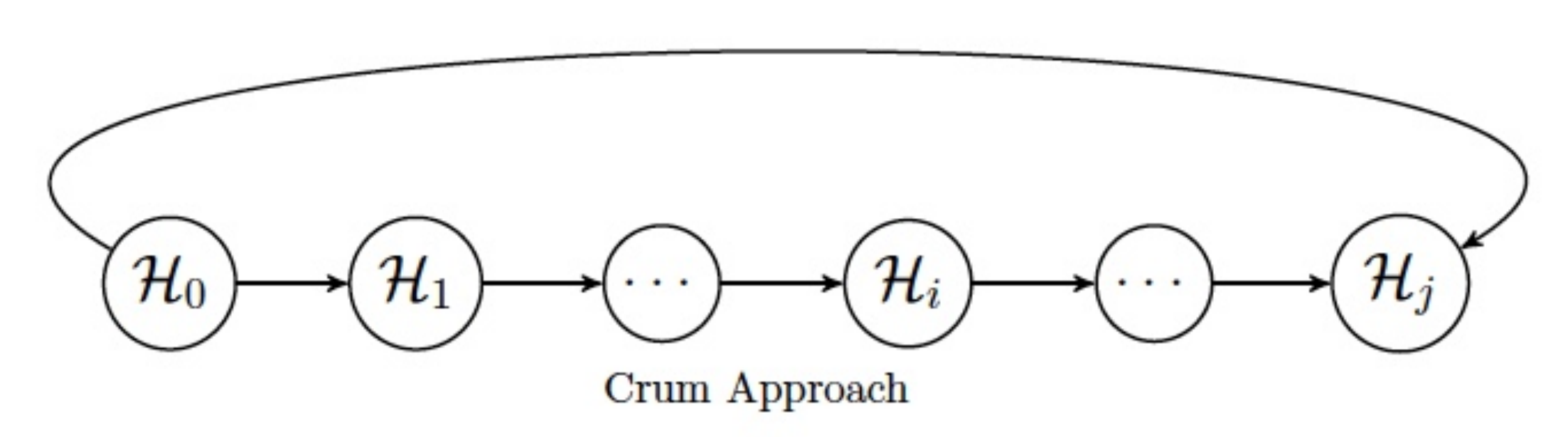}
\end{center}
This approach can be illustrated as follows:  consider the eigenfunctions $\phi_{1,n}(x),~n=1,2,\dots$ of the Hamiltonian $\mathcal H_1$ generated using $\phi_{0,n}(x),n=0,1,2,\dots$; the solutions of an initial Hamiltonian $\mathcal H_0=-\partial_{xx}+V(x)$; and employed in a second Darboux transformation to obtain
\begin{align}\label{eq43}
\phi_{2,n}(x)
&=\left(\dfrac{d}{dx}-\dfrac{d}{dx}\log({\phi_{1,1}(x)})\right)\,\phi_{1,n}(x)=\dfrac{W(\phi_{0,0}(x),\phi_{0,1}(x),\phi_{0,n}(x))}{W(\phi_{0,0}(x),\phi_{0,1}(x))},\quad n=2,3,\dots
\end{align}
where we perform the computations using \eqref{eq39} and the definition of the Wronskian.
 This approach, in turn, generates a new class of solvable Schr\"odinger equation
\begin{align}\label{eq44}
-\dfrac{d^2}{dx^2}\phi_{2,n}(x)&+\left(V(x)- 2\dfrac{d^2}{dx^2}\bigg(\log\bigg[W(\phi_{0,0}(x),\phi_{0,1}(x))\bigg]\bigg)\right)\phi_{2,n}(x)=E_n\phi_{2,n}(x),\quad n=2,3,\dots.
\end{align}
expressed entirely in terms of the eigenfunctions of the Hamiltonian $\mathcal H_0$ with no reference to the intermediate Hamiltonian $\mathcal H_1$. For  the eigenvalue problem \eqref{eq44}, we may now employ the third transformation
\begin{align}\label{eq45}
\phi_{3,n}(x)=\left(\dfrac{d}{dx}-\dfrac{d}{dx}\log \phi_{2,2}(x)\right)\,\phi_{2,n}(x)=\dfrac{W(\phi_{0,0}(x),\phi_{0,1}(x),\psi_{0,2}(x),\phi_{0,n}(x))}{W(\phi_{0,0}(x),\phi_{0,1}(x),\phi_{0,2}(x))},\quad n=3,4,\dots
\end{align}
to obtain the following solvable class of Schr\"odinger equations
\begin{align}\label{eq46}
-\dfrac{d^2}{dx^2}\phi_{3,n}(x)+\left(V(x)-2\dfrac{d^2}{dx^2}\bigg(\log\bigg[W(\phi_{0,0}(x),\phi_{0,1}(x),\phi_{0,2}(x)\bigg]\bigg)
\right)\phi_{3,n}(x)=E_n \phi_{3,n}(x),\quad n=3,4,\dots.
\end{align}
without reference to the intermediate Hamiltonians $\mathcal H_1$ and $\mathcal H_2$.   This process may be generalized to include the case of $j$-times repeated Darboux transformation, expressed entirely in terms of the eigenfunctions for the initial Hamiltonian $\mathcal H_0$ to give: \emph{The transformed functions
\begin{align}\label{eq47}
\psi_{j,n}(x)&=\left(\dfrac{d}{dx}-\dfrac{d}{dx}\log{\psi_{j-1,j-1}(x)}\right)\,\psi_{j-1,n}(x)\notag\\
&=\dfrac{W(\phi_{0,0}(x),\phi_{0,1}(x),\phi_{0,2}(x),\dots,\phi_{0,k-1}(x),\phi_{0,n}(x))}{W(\phi_{0,0}(x),\phi_{0,1}(x),\phi_{0,2}(x),\dots,\phi_{0,k-1}(x))},\quad
n=k,k+1,k+2,\dots
\end{align}
satisfy the Schr\"odinger equation
\begin{align}\label{eq48}
-\dfrac{d^2}{dx^2}\phi_{j,n}(x)+\left(V(x)-2\dfrac{d^2}{dx^2}\left[\log \bigg(W(\phi_{0,0}(x),\phi_{0,1}(x),\phi_{0,2},\dots,\phi_{0,j-1}(x)\right)\bigg]\right)\phi_{j,n}(x)=E_n\phi_{j,n}(x),
\end{align}
where $n=j,j+1,j+2,\dots.$ The proof of Curm's formulation \cite{crum} is based on the following Wronski identity:
\begin{align}\label{eq49}
W(\phi_{0,0},\phi_{0,1},\dots,\phi_{0,j-1})&W(\phi_{0,0},\phi_{0,1},\dots,\phi_{0,j},\phi_{0,n})\notag\\
&=W(W(\phi_{0,0},\phi_{0,1},\dots,\phi_{0,j}),W(\phi_{0,0},\phi_{0,1},\dots,\psi_{0,j-1},\phi_{0,n})).
\end{align}
The general expression of the normalization constants is given in terms of the normalization of $\mathcal H_0$ as:
\begin{equation}\label{50}
\int_a^b \phi_{j,n}(x)\phi_{j,m}(x) dx
=\,\mu_n\,\delta_{nm}\prod_{i=0}^{j-1}(E_n-E_i),\quad where\quad
\mu_n=\int_a^b [\phi_{0,n}(x)]^2 dx.
\end{equation}
}
\section{The Crum-Darboux transformation and the generalized Hulth\'en potential}
\no From section \ref{sec3}, we may write  the exact solutions of the Schr\"odinger equation
\begin{align}\label{eq51}
-\dfrac{d^2\psi_{0,n}(r)}{dr^2}-\dfrac{v\,e^{-r}}{1-q\,e^{-r}}\,\psi_{0,n}(r)=\,\mathscr E_n\,\psi_{0,n}(r),\qquad r\in[\log(q),\infty),\quad q>0,\quad \psi_{0,n}(\log(q))=\psi_{0,n}(\infty)=0,\end{align}
are simply, for $n=0,1,2,\dots$,
\begin{align}\label{eq52}
\mathscr E_n&=-\left(\dfrac{v}{2\,q\,(n+1)}-\dfrac{n+1}{2}\right)^2,\notag\\
\psi_{0,n}(r)&=N_n \left(1-q e^{-r}\right) e^{-\left(\frac{v}{2 (n+1) q}-\frac{n+1}{2}\right)\,r} \, _2F_1\left(-n,\frac{v}{(n+1)q}+1;\frac{v}{(n +1)q}-n;q\, e^{-r}\right),
\end{align}
up to the normalization constant evaluated using the following definite integral (see equation \eqref{eq28} and \eqref{eq29}):
\begin{align}\label{eq53}
N_n^2&\int_{\log(q)}^\infty (1-q\, e^{-r})^{2}\, e^{-\left(\frac{v}{q (n+1)}-n-1\right)\, r} \left[\, _2F_1\left(-n,\frac{v}{q(1+n)}+1; \frac{v}{q(1 + n )}-n;q\, e^{-r}\right)\right]^2dr=1
\end{align}
as
\begin{align}\label{eq54}
&N_n=\left\{
\frac{2 q^{n+1-\frac{v}{q\, (n+1)}}\, \Gamma \left(\frac{v}{q\,  (n+1)}-n-1\right)}{\Gamma \left(\frac{v}{q\,  (n+1)}-n+2\right)}\bigg[\frac{q\, (n+1)\left(v-q\,  (n+1)^2\right) \left(\frac{v}{q\,  (n+1)}+1\right)_n}{\left(v-q\,  (n+1)\right)
\left(2\,q\,  (n+1)+v\right) \left(-\frac{v}{q\,  (n+1)}-1\right)_n}\right.\notag\\
&\times\left.{}_3F_2\left(\begin{array}{lll}1-n,&\frac{v}{q\,  (n+1)}+1,&\frac{v}{q\,  (n+1)}\\
\frac{v}{q\,  (n+1)}+3,&\frac{v}{q\,  (n+1)}-n,& \end{array}\bigg|1\right)+\, _3F_2\left(\begin{array}{lll}-n,&\frac{v}{q\,  (n+1)}+1,&\frac{v}{q\, (n+1)}-n-1\\
\frac{v}{q\,  (n+1)}-n,&\frac{v}{q\, (n+1)}-n+2& \end{array}\bigg|1\right)\bigg]\right\}^{-1/2},
\end{align}
and subject to the parameter constraints
$$ v>0,\quad 0< q<v,\quad 0\leq n<\sqrt{\frac{v}{q}}-1,$$
where $n$ is an integer.
\subsection{First Transformation}
\noindent Consider the Darboux transformation
\begin{align}\label{eq55}
\psi_{1,n}(r) &= \left[\dfrac{d}{dr} - \dfrac{1}{\psi_{0,0}(r)} \dfrac{d\psi_{0,0}(r)}{dr}\right] \psi_{0,n}(r)\equiv \dfrac{W(\psi_{0,0}(r),\psi_{0,n}(r))}{\psi_{0,0}(r)} \, ,\quad n=1,2,\dots \, ,
\end{align}
where the seed function $\psi_{0,0}(r)$ given as the ground-state wave-function \eqref{eq52}, from which a new solvable potential is obtained:
\begin{align}\label{eq56}
V_{1}(r) = V(r) - 2 \dfrac{d^{2}}{dr^{2}} \log\psi_{0,0}(r) &=-\frac{v\,e^{-r}}{1-q\,e^{-r}}+ \frac{2\, q\, e^{-r}}{\left(1-q\,e^{-r}\right)^2}
\end{align}
Thus, the Schr\"odinger equation with the potential $V_1$ is exactly solvable
\begin{align}\label{eq57}
- \dfrac{d^{2}}{dr^{2}} \psi_{1,n}(r) + \left(-\frac{ve^{-r}}{1-qe^{-r}}+ \frac{2\, q\, e^{-r}}{\left(1-qe^{-r}\right)^2}\right)\psi_{1,n}(r) &=- \,\mathscr E_n\psi_{1,n}(r) \, , \quad n = 1, 2, \dots \, ,
\end{align}
with (up to normalization constant) exact wave function solutions, for $n=1,2,\dots$,
\begin{align}\label{eq58}
\psi_{1,n}(r)&= \frac{n \left(e^r-q\right)}{2q (n+1)}e^{-\frac{1}{2} r \left(\frac{v}{n q+q}-n+1\right)}\left[2q (n+1) \, _2F_1\left(1-n,\frac{v}{n q+q}+1;\frac{v}{n q+q}-n;q\,e^{-r} \right) \right. \notag\\
&
\left.+(v-(n +1)q) \, _2F_1\left(-n,\frac{v}{n q+q}+1;\frac{v}{n q+q}-n;q\,e^{-r}\right)\right],\end{align}
subject to $v > 0, q > 0, 1\le n < \sqrt{(q + v)/q} $. Using the contiguous relation
\begin{align}\label{eq59}
(\gamma-\alpha-\beta){}_2F_1(\alpha,\beta;\gamma;z)+\alpha(1-z)\,{}_2F_1(\alpha+1,\beta;\gamma;z)-(\gamma-\beta)\,{}_2F_1(\alpha,\beta-1;\gamma;z)=0,
\end{align}
equation \eqref{eq58} can be written in more compact form as:
\begin{align}\label{eq60}
\psi_{1,n}(r)&=\frac{n\,  (q\,n+q+v)}{2 (n+1) q}\, e^{-\frac{r}{2} \left(\frac{v}{n q+q}-n-1\right)}\left(1-qe^{-r}\right)^2\, _2F_1\left(1-n,\frac{v}{n q+q}+2;\frac{v}{n q+q}-n;e^{-r} q\right),\notag\\
&\hskip4true in(n=1,2,\dots,\mathfrak{n}<\sqrt{(q + v)/q}).
\end{align}
\subsection{Second transformation}
\no Using these exact solutions \eqref{eq60}, it is possible via Crum's approach to generate sequential transformation of the Hulth\'{e}n potential with
\begin{align}\label{eq61}
\psi_{2,n}(r) &= \left[\dfrac{d}{dr} - \dfrac{\psi_{1,1}'(x)}{\psi_{1;1}(r)}\right] \psi_{1,n}(r)\equiv \frac{W\left(\psi _{0,0}(r),\psi _{0,1}(r),\psi _{0,n}(r)\right)}{W\left(\psi _{0,0}(r),\psi _{0,1}(r)\right)} \, , \quad n=2,3,\dots \, ,
\end{align}
where the seed function $\psi_{1;1}(r)$ given by the ground-state wave function \eqref{eq60} for $n=1$. Thus, a new solvable potential is obtained:
\begin{align}\label{eq62}
V_2(r) &= V_{1}(r)-2\dfrac{d^2}{dr^2}\log\big[\psi_{1;1}(r)\big] =V(r)-2\,\dfrac{d^2}{dr^2}\log W(\psi_{0,0}(r),\psi_{0,1}(r))=-\frac{v\,e^{-r}}{1-q\,e^{-r}}
+\frac{6\, q\, e^{-r}}{\left(1-q\,e^{-r}\right)^2} \, .
\end{align}
This potential has exact analytic solutions of the Schr\"odinger equation
\begin{equation}\label{eq63}
- \dfrac{d^{2}}{dr^{2}} \psi_{2,n}(r)+\left(-\frac{v\,e^{-r}}{1-q\,e^{-r}}+\frac{6\, q\, e^{-r}}{\left(1-q\,e^{-r}\right)^2}\right)\psi_{2,n}(r) = - \,\mathscr E_n\psi_{2,n}(r) \, , \quad n = 2, 3, \dots \, ,
\end{equation}
with exact solutions (up to normalization constant) given for $n=2,3,\dots,$
\begin{align}\label{eq64}
&\psi_{2,n}(r)= \frac{n(n-1)(v+q\,(n+1))(v+2\, q\,(n+1))}{8 (1+n)^2 q^2}e^{-\frac{r}{2} \left(\frac{v}{q+n q}-n-1\right)} (1-q\,e^{-r})^2\notag\\
&\hskip0.5true in\times\left(\, _2F_1\left(1-n,\frac{v}{n q+q}+2;\frac{v}{n q+q}-n;e^{-r} q\right)\right.\notag\\
&\hskip1true in\left.-\frac{4 \,q^2\, (n+1)}{\left(n q+n^2 q-v\right)}\,e^{-r} \, _2F_1\left(2-n,\frac{v}{n q+q}+3;\frac{v}{n q+q}-n+1;e^{-r} q\right)\right)
\end{align}
That can be written as
\begin{align}\label{eq65}
\psi_{n;2}(r)&= \frac{n(n-1)(v+(n+1) q) (v+2 (n+1) q) }{8\,q^2 (n+1)^2}\,(1-q e^{-r})^3\, e^{-\frac{r}{2}  \left(\frac{v}{n q+q}-n-1\right)}\notag\\
&\times \, _2F_1\left(2-n,\frac{v}{n q+q}+3;\frac{v}{n q+q}-n;e^{-r} q\right),
\end{align}
subject to $v > 0, ~0 < q < v/3, ~2 \le  n < \sqrt{(q + v)/q}.$

\subsection{The $j^{\text{th}}$ Transformation}
\no The class of generalized Hulth\'{e}n potentials given by an arbitrary $j^{\text{th}}$ transformation, $j=1,2,\dots$ can be established using Crum's approach, see Section \ref{sec3}, to give
\begin{align}\label{eq66}
V_{j}(r) &\equiv -\frac{v\, e^{-r}}{1-q\,e^{-r}} - 2 \dfrac{d^{2}}{dr^{2}} \log W\big(\psi_{0;0}(r),\psi_{1;0}(r),\dots,\psi_{j-1;0}(r)\big),\quad j=1,2,\dots,\mathfrak n.
\end{align}
We shall show, by induction on $j$, that
\begin{align}\label{eq67}
- 2 \dfrac{d^{2}}{dr^{2}} \log W\big(\psi_{0;0}(r),\psi_{1;0}(r),\dots,\psi_{j-1;0}(r)\big)=\dfrac{ j(j+1)\, q\,e^{r}}{(e^r-q)^2}
\end{align}
whence
\begin{equation}\label{eq68}
V_j(r)=-\frac{ve^{-r}}{1-q\,e^{-r}}+\dfrac{ j(j+1)\, q\,e^{r}}{(e^r-q)^2},\qquad j=1,2,\cdots
\end{equation}
For $j=1$, using
$W\big(\psi_{0;0}(r))=\psi_{0;0}(r)$
we note for
\begin{align}\label{eq69}
\psi_{0,0}(r)&=e^{-r \left(\frac{v}{2q}-\frac12\right)} \left(1-q\,e^{-r}\right),\qquad  v > q>0,\notag\\
\dfrac{d}{dr}\log \psi_{0;0}(r)&=\frac{1}{2}+\frac{q}{e^r-q}-\frac{v}{2 q},\qquad \dfrac{d^2}{dr^2}\log \psi_{0;0}(r)=-\frac{q\,e^r}{\left(e^r-q\right)^2}.
\end{align}
and equation \eqref{eq67} is true for $j=1$. Assume, equation \eqref{eq67} is true for $j=j'$, that is to say
\begin{align}\label{eq70}
- 2 \dfrac{d^{2}}{dr^{2}} \log W\big(\psi_{0;0}(r),\psi_{1;0}(r),\dots,\psi_{j'-1;0}(r)\big)=\dfrac{ j'(j'+1)\, q\,e^{r}}{(e^r-q)^2}
\end{align}
then for
 $j=j'+1$, since
\begin{align}\label{eq71}
- 2 \dfrac{d^{2}}{dr^{2}} \log\, &W\big(\psi_{0;0}(r),\psi_{1;0}(r),\dots,\psi_{j'-1;0}(r),\psi_{j';0}(r)\big)\notag\\
=&- 2 \dfrac{d^{2}}{dr^{2}} \log W\big(\psi_{0;0}(r),\psi_{1;0}(r),\dots,\psi_{j'-1;0}(r)\big)- 2 \dfrac{d^{2}}{dr^{2}} \log \frac{W\big(\psi_{0;0}(r),\psi_{1;0}(r),\dots,\psi_{j'-1;0}(r),\psi_{j';0}(r)\big)}{W\big(\psi_{0;0}(r),\psi_{1;0}(r),\dots,\psi_{j'-1;0}(r)\big)}\notag\\
&=\dfrac{ j'(j'+1)\, q\,e^{r}}{(e^r-q)^2}- 2 \dfrac{d^{2}}{dr^{2}} \log \psi_{j';j'}(x)=\dfrac{ j'(j'+1)\, q\,e^{r}}{(e^r-q)^2}+V_{j'+1}(r)-V_{j'}(r)\notag\\
&=\dfrac{v\,e^{-r}}{1-q\,e^{-r}}+V_{j'+1}(r),
\end{align}
which ensures the truth of the identity \eqref{eq67}. This potential has exact analytic solutions
\begin{align}\label{eq72}
\psi_{j,n}(r) &=\psi_{j-1,j}'(r)-\dfrac{\Psi_{j-1,j-1}'(r)}{\psi_{j-1,j-1}(r)}\psi_{j-1,n}(r)
= \dfrac{W\big(\psi_{0;0}(r),\psi_{1;0}(r),\dots,\psi_{j-1;0}(r),\psi_{n;0}(r)\big)}{W\big(\psi_{0;0}(r),\psi_{1;0}(r),\dots,\psi_{j-1;0}(r)\big)},~n=j,j+1,\dots \, \mathfrak n,
\end{align}
where $\mathfrak n$ indicate the finiteness of the discrete bound states. The eigenfunctions \eqref{eq69} are the solutions of the Schr\"odinger equation
\begin{align}\label{eq73}
-\dfrac{d^2\psi_{n;j}(r)}{dr^2} +\left(\dfrac{ j(j+1)\, q\,e^{-r}}{(1-qe^{-r})^2}-\frac{v\,e^{-r}}{1-q\,e^{-r}}\right)\Psi_{n;j}(r)=- \,\mathscr E_n\,\Psi_{n;j}(r),\qquad r\in [\log(q),\infty)
\end{align}
with
\begin{align}\label{eq74}
\mathscr E_n&=-\left(\dfrac{v}{2\,q\,(n+1)}-\dfrac{n+1}{2}\right)^2,\quad n=j,~j+1,\dots
\end{align}
subject to $v > 0, q > 0 , j \leq  n < \sqrt{(q + v)/q}.$ Obviously, the evaluation of the general expression for $\psi_{j,n}(r)$ using the relation \eqref{eq69} is not straightforward.   However, we can find a general expression by analyzing the possible solutions  of the Schr\"odinger equation
\begin{align}\label{eq75}
-\dfrac{d^2\psi(r)}{dr^2} +\left(\dfrac{\mu \,e^{-r}}{(1-q\, e^{-r})^2}-\frac{v\,e^{-r}}{1-q\,e^{-r}}\right)\psi (r)=\mathfrak E_n\,\psi(r),\qquad r\in [\log(q),\infty)
\end{align}
where $\mu$ is an arbitrary constant that supports the existence of discrete bound-states.
\section{Extended Hulth\'en's potential: exact solutions}
\noindent In this section, we analyze the exact solutions of equation \eqref{eq75} which will allows us to obtain a compact formula for the $j$-transformed $\psi_{j,n}(r)$ as given by \eqref{eq69}. Using a similar approach to that discussed in Section \ref{sec2}, it is not difficult to show that the change of variable $z=e^{-r}$ along with the analysis of the regular singular points implies by means of the ansatz solution
\begin{align}\label{eq76}
 \psi(z)=z^\eta\left(1-q\,z\right)^s f(z),\quad where\quad (\eta^2+\mathfrak E_n=0, \qquad qs^2-qs-\mu=0).
\end{align}
the following second-order differential equation for $f(z)$
\begin{equation}\label{eq77}
z\,(1-q\,z)\,f''(z)+\left[1+2\, \eta-q\, (1+2 \eta+2 s) z\right]f'(z)+\left[v-q \left(2 \eta s+s^2\right)\right]f(z)=0
\end{equation}
The exact solutions of this equation are given in terms of the Gauss hypergeometric functions as
\begin{align}\label{eq78}
f(z)&={} _2F_1\left(\eta+s-\sqrt{\eta^2 +\frac{v}{q}},\eta+s+
\sqrt{\eta^2+\frac{v}{q}};1+2\, \eta;q\, z\right),
\end{align}
up to the normalization constant where $\eta=\sqrt{-\mathfrak E_n}$ and $s_{\scriptscriptstyle \pm}=\frac{1}{2}\pm\sqrt{\frac{\mu}{q}+\frac{1}{4}}.$ Imposing the termination condition on the hypergeometric function, to obtain polynomial solutions, implies the exact solutions of Schrodinger's equation \eqref{eq75} as
\begin{align}\label{eq79}
\psi(r)&=e^{-\left(\frac{v}{2 q (n+s_{\scriptscriptstyle +})}-\frac{n+s_{\scriptscriptstyle +}}{2}\right)\,r} \,\left(1-q\,e^{-r}\right)^{s_{\scriptscriptstyle +}}\, {} _2F_1\left(-n,s_{\scriptscriptstyle +}+\frac{v}{q (n+s_{\scriptscriptstyle +})};1-n-s_{\scriptscriptstyle +}+\frac{v}{q (n+s_{\scriptscriptstyle +})};q\, \,e^{-r}\right),
\end{align}
where
\begin{align}\label{eq80}
\eta=\sqrt{-\mathfrak E_n},\qquad s_{\scriptscriptstyle +} =  \frac{1}{2}+\sqrt{\frac{\mu}{q}+\frac{1}{4}},\qquad \mathfrak E_n= -\left(\frac{v}{2\,q\,(n+s_{\scriptscriptstyle +})}-\frac{n+s_{\scriptscriptstyle +}}{2}\right)^2,
\end{align}
up to the normalization constant that, as before, can be evaluated exactly. With $\mu=q\,j\,(j+1)$, we obtain $s_{\scriptscriptstyle +}=j+1, j=0,1,2,\dots,$ which proves the consistency between $\mathfrak E_n$ and $\mathcal E_n$ as given by \eqref{eq74}. Finally, we can now write, for
$q < (2 v)/(1 + 2 j + j^2),
 0 \leq  n < -1 - j + \sqrt{2v/q}$,
 \begin{align}\label{eq81}
\psi_{j,n}(r) &= \dfrac{W\big(\psi_{0;0}(r),\psi_{1;0}(r),\dots,\psi_{j-1;0}(r),\psi_{n;0}(r)\big)}{W\big(\psi_{0;0}(r),\psi_{1;0}(r),\dots,\psi_{j-1;0}(r)\big)}\notag\\
&=e^{-\left(\frac{v}{2 q (n+j+1)}-\frac{n+j+1}{2}\right)\,r} \,\left(1-q\,e^{-r}\right)^{j+1}\, {} _2F_1\left(-n,j+1+\frac{v}{q (n+j+1)};\frac{v}{q (n+j+1)}-n-j;q\, \,e^{-r}\right),\notag\\
&=e^{-\left(\frac{v}{2 q (n+1)}-\frac{n+1}{2}\right)\,r} \,\left(1-q\,e^{-r}\right)^{j+1}\, {} _2F_1\left(j-n,j+1+\frac{v}{q (n+1)};\frac{v}{q (n+1)}-n;q\, \,e^{-r}\right),~n=j,j+1,\dots
\end{align}
which is total agreement with the normalization constant obtain through the identity
\begin{equation}\label{82}
\int_{\log q}^\infty \psi_{j,n}(r)\psi_{j,m}(r) dx
=\frac{(-j)_j(j+2 n+2)_j }{(2^j(n+1)_j)^2}\left(n-\frac{v}{(j+n+1) q}+1\right)_j \left(n+\frac{v}{(j+n+1) q}+1\right)_j\times N_{n}\,\delta_{nm}\end{equation}
and $N_{n}$ is given by equation \eqref{eq54}.
\section{Conclusion}

\noindent General expressions for the energy eigenvalues and wave function solutions are obtained for Schr\"odinger's equation with the generalized Hulth\'en potential.  The simplified closed--form expressions for the normalization constants for arbitrary $q>0$ in terms of the generalized Hypergeometric functions ${}_3F_2$ with the unit argument are new results. These include, as a particular case, the closed--form expression for the normalization constants of the classical Hulth\'en potential $q=1$. It is also of interest to note that the double sum in equation \eqref{eq17} can be evaluated in terms of the terminating generalized Kamp\'e de F\'eriet function with unit arguments to yield
\begin{align}\label{eq83}
&I_{nm}=\int_{\log(q)/\delta}^\infty e^{-\frac{(2+m+n) \left(2 \mu-q\,\delta^2 (1+m) (1+n)\right) x}{2 \delta (1+m) (1+n) q}} \left(1-q\,e^{-\delta x}\right)^2\, _2F_1\left(-n,\frac{2 \mu}{\delta^2 (n+1) q}+1;\frac{2 \mu}{\delta^2 (n+1) q}-n;q e^{-\delta\, x}\right)\notag\\
&\times \, _2F_1\left(-m,\frac{2 \mu}{\delta^2 (m+1) q}+1;\frac{2 \mu}{\delta^2 (m+1) q}-m;q e^{-\delta\, x}\right)dx\notag\\
&=\frac{2 q^{1+\frac{m+n}{2}-\frac{\mu (m+n+2)}{\delta^2 (m+1) (n+1) q}} \Gamma \left(\frac{\mu (m+n+2)}{\delta^2 (m+1) (n+1) q}-\frac{m+n}{2}-1\right)}{\delta\, \Gamma \left(2-\frac{m+n}{2}+\frac{\mu (m+n+2)}{\delta^2 (m+1) (n+1) q}\right)}\notag\\
&\times F_{1:1;1}^{1:2,2}\left[\begin{array}{lll}
\frac{\mu (m+n+2)}{\delta^2 (m+1) (n+1) q}-\frac{m+n}{2}-1:&-n,\quad\frac{2 \mu}{q \delta^2 (n+1)}+1;&-m, \quad \frac{2 \mu}{q \delta^2 (m+1)}+1;\\ \\
\frac{\mu (m+n+2)}{\delta^2 (m+1) (n+1) q}-\frac{m+n}{2}+2:&\frac{2 \mu}{q \delta^2 (n+1)}-n; &\frac{2 \mu}{q \delta^2 (m+1)}-m;\end{array}\bigg|  1,1
\right]\end{align}
Thus we obtain as a byproduct a simplified expression for the terminating generalized Kamp\'e de F\'eriet function. It worth mentioning some of the earlier approaches can also be used to study the deformed Hulth\'en potential, for example, the improved quantization rule \cite{Dong2007c}.

\section{Acknowledgments}
\medskip
\noindent Partial financial support of this work under Grant Nos. GP3438 and GP249507 from the
Natural Sciences and Engineering Research Council of Canada
 is gratefully acknowledged by us (respectively RLH and NS).
\section*{Appendix I: Normalization relation}
\noindent To prove the normalization \eqref{eq42}, we note first
\begin{align*}
\int_a^b \psi_n(x,E_n)\psi_m(x,E_m) dx&=\int_a^b \phi_n'(x,E_n)\phi_m'(x,E_m)dx-\int_a^b \phi_m'(x,E_m)\phi_n(x,E_n)\dfrac{d}{dx}\log{\phi_0(x,\lambda_0)}dx\\
&-\int_a^b \phi_n'(x,E_n)\,\phi_m(x,E_n))\dfrac{d}{dx}\log{\phi_0(x,\lambda_0)}dx+\int_a^b \left(\dfrac{\phi_0'(x,\lambda_0)}{\phi_0(x,\lambda_0)}\right)^2\,\phi_n(x,E_n)\,\phi_m(x,E_n)dx.
\end{align*}
Using integration by parts and the boundary conditions, it is not difficult to show that
\begin{align*}
\int_a^b \phi_n'(x,E_n)&\phi_m'(x,E_m)dx=E_n\mu_{n}\delta_{nm}- \int_a^b  \phi_m(x,E_m)V(x)\phi_n(x,E_n)dx\\
\int_a^b \phi_m'(x,E_m)&\phi_n(x,E_n)\dfrac{d}{dx}\log{\phi_0(x,E_0)}dx
=\phi_n(x,E_n)\phi_m(x,E_m))\dfrac{d}{dx}\log{\phi_0(x,E_0)}\bigg|_a^b\\
&-
\int_a^b \phi_n(x,E_n)\phi_m'(x,E_m)\dfrac{d}{dx}\log{\phi_0(x,E_0)}dx-\int_a^b \phi_n(x,E_n)\phi_m(x,E_m)\dfrac{d^2}{dx^2}\log{\phi_0(x,E_0)}dx\\
\int_a^b \phi_n'(x,\lambda_n)&\phi_m(x,\lambda_m)\dfrac{d}{dx}\log{\phi_0(x,\lambda_0)}dx=
\phi_n(x,E_n)\phi_m(x,E_m))\dfrac{d}{dx}\log{\phi_0(x,E_0)}\bigg|_a^b\\
&-
\int_a^b \phi_m(x,E_m)\,\phi_n'(x,E_n)\dfrac{d}{dx}\log{\phi_0(x,E_0)}dx-\int_a^b \phi_n(x,E_n)\phi_m(x,E_m)\dfrac{d^2}{dx^2}\log{\phi_0(x,E_0)}dx\\
\int_a^b \left(\dfrac{\phi_0'(x,\lambda_0)}{\phi_0(x,\lambda_0)}\right)^2&\,\phi_n(x,E_n)\,\phi_m(x,E_m)dx=\int_a^b \left(V(x)-E_0-\dfrac{d^2}{dx^2}\log\phi_0(x,\lambda_0)\right)\,\phi_n(x,\lambda_n)\,\phi_m(x,\lambda_m)dx,
\end{align*}
from which the assertion \eqref{eq42} is proved.

\end{document}